\documentclass[letter,twocolumn,aps, showpacs]{revtex4}

\usepackage{graphicx}
\usepackage{dcolumn}
\usepackage{bm}

\newcommand{\pderiv}[2]{\frac{\partial #1}{\partial #2}}
\newcommand{\pb}{p_{\rm b}}
\newcommand{\pd}{p_{\rm d}}
\newcommand{\pdc}{p_{\rm {d_c}}}
\newcommand{\rhobar}{\bar\rho}
\newcommand{\Bi}{\mathcal{B}_{\rm i}}
\newcommand{\Di}{\mathcal{D}_{\rm i}}

\begin{document}

\preprint{APS/123-QED}

\title{Cluster geometry and survival probability in systems driven by reaction-diffusion
dynamics}

\author{Alastair L Windus}
\author{Henrik Jeldtoft Jensen}
 \email{h.jensen@imperial.ac.uk}
\affiliation{The Institute for Mathematical Sciences. 53 Prince's Gate, South Kensington, London. SW7 2PG.}
\affiliation{Department of Mathematics, Imperial College London, South Kensington Campus, London. SW7 2AZ.}

\date{\today}

\begin{abstract}
We consider a reaction-diffusion model incorporating the reactions $A\rightarrow\phi$,
$A\rightarrow2A$ and $2A\rightarrow 3A$. Depending on the relative rates
for sexual and asexual reproduction of the quantity $A$, the model exhibits
either a continuous or first-order absorbing phase transition to an extinct
state. A tricritical point separates the two phase lines. As well as briefly
examining this critical behavior in 2+1 dimensions, we pay particular attention
to the cluster geometry. We observe the different cluster structures that
form at criticality for the three different types of critical behavior and
show that there exists a linear relationship for the probability of survival against initial cluster size at the tricritical point only.
\end{abstract}


\maketitle

Being able to identify the important geometrical structures is known to greatly facilitate the understanding of the underlying physical mechanism \cite{Nelson_Defects}. In this letter we describe how the tricritical point in a certain class of nonequilibrium population models is characterized by the unique structure of the spatial cluster. Nonequilibrium phase transitions are of great relevance to disciplines as wide-ranging as traffic flow \cite{Chowdhury}, chemical
reactions \cite{Schlogl_Chemical} and atmospheric studies \cite{Jin}. Of particular interest is
the idea of universality, where different models display identical scaling
functions and critical exponents close to the critical point. By far the
most well-known and studied universality class out of equilibrium is directed
percolation (DP). The robustness of DP led Janssen and Grassberger \cite{Grassberger_On,Janssen_Non}
to the conjecture that all models with a scalar order parameter that exhibit a continuous phase transition from an active to a single absorbing state belong to the class. A universality class closely associated with DP is tricritical directed percolation (TDP), which has recently been studied both from steady-state \cite{Lubeck_Tricritical} and dynamical \cite{Grassberger_Tricritical} simulations.

The process of TDP incorporates higher-order terms than DP, with the multicritical
behavior occurring if the lower-order reactions vanish on a coarse-grained
scale \cite{Ohtsuki2}. L\"ubeck \cite{Lubeck_Tricritical} examined tricritical behavior by adding the pair reaction $2A\rightarrow 3A$ to the contact process \cite{Harris} while Grassberger \cite{Grassberger_Tricritical} used a generalisation of the Domany-Kinzel model \cite{Domany}
in 2+1 dimensions.

Recently, we examined a simple reaction-diffusion model with the reactions
$A \rightarrow \phi$ and $2A\rightarrow 3A$ \cite{Windus}, which
exhibited a phase transition that is continuous, and DP in 1+1 dimensions
and first first-order in higher dimensions. Although, from the mean field, the first-order phase transition was expected in all dimensions, the larger
fluctuations in the (1+1)-dimensional case are likely to destabilize
the ordered phase \cite{Marro}, resulting in the observed DP transition.

\begin{figure}[t]
        \centering\noindent
        \includegraphics[width=2cm,height=2cm]{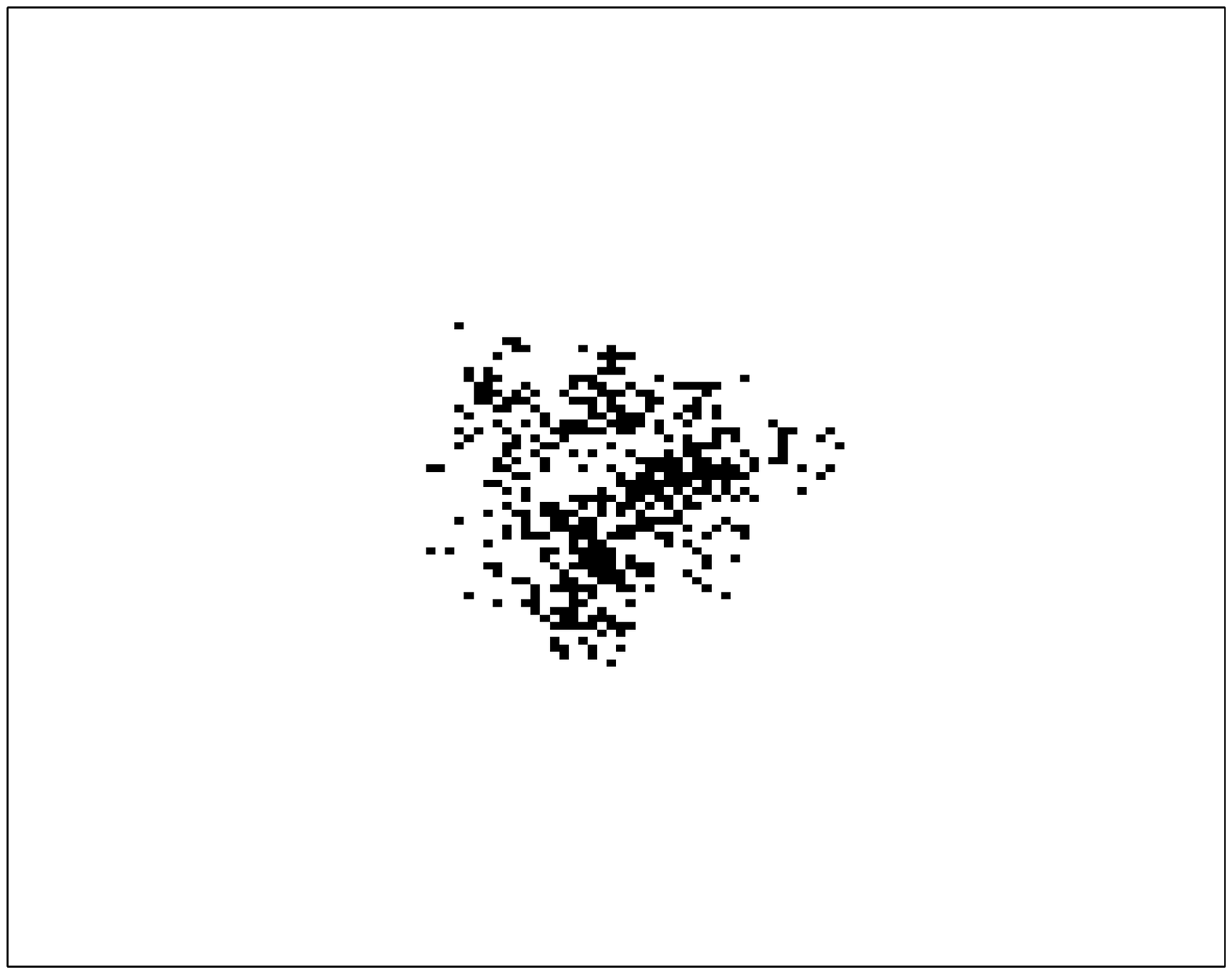}
        \includegraphics[width=2cm,height=2cm]{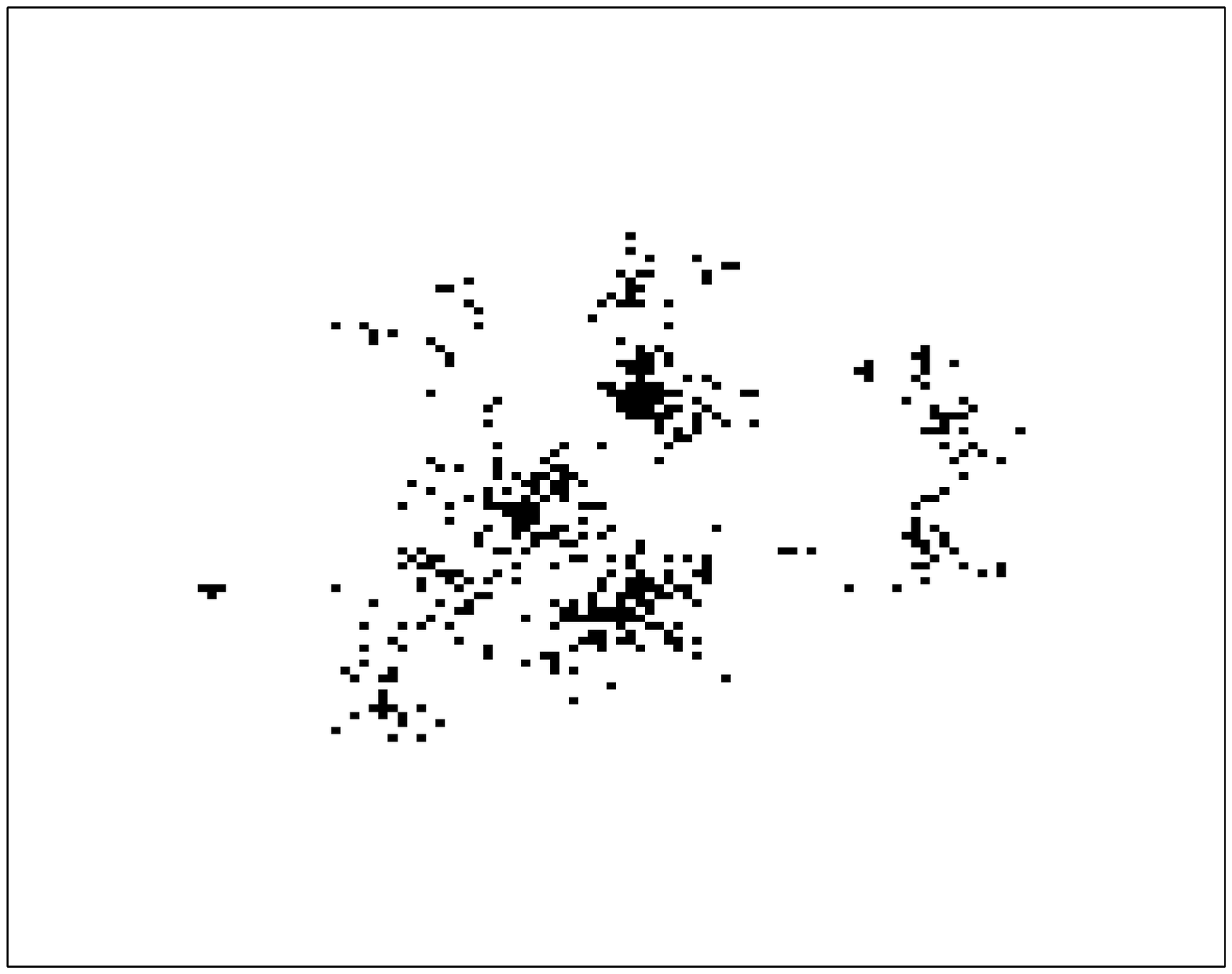}
        \includegraphics[width=2cm,height=2cm]{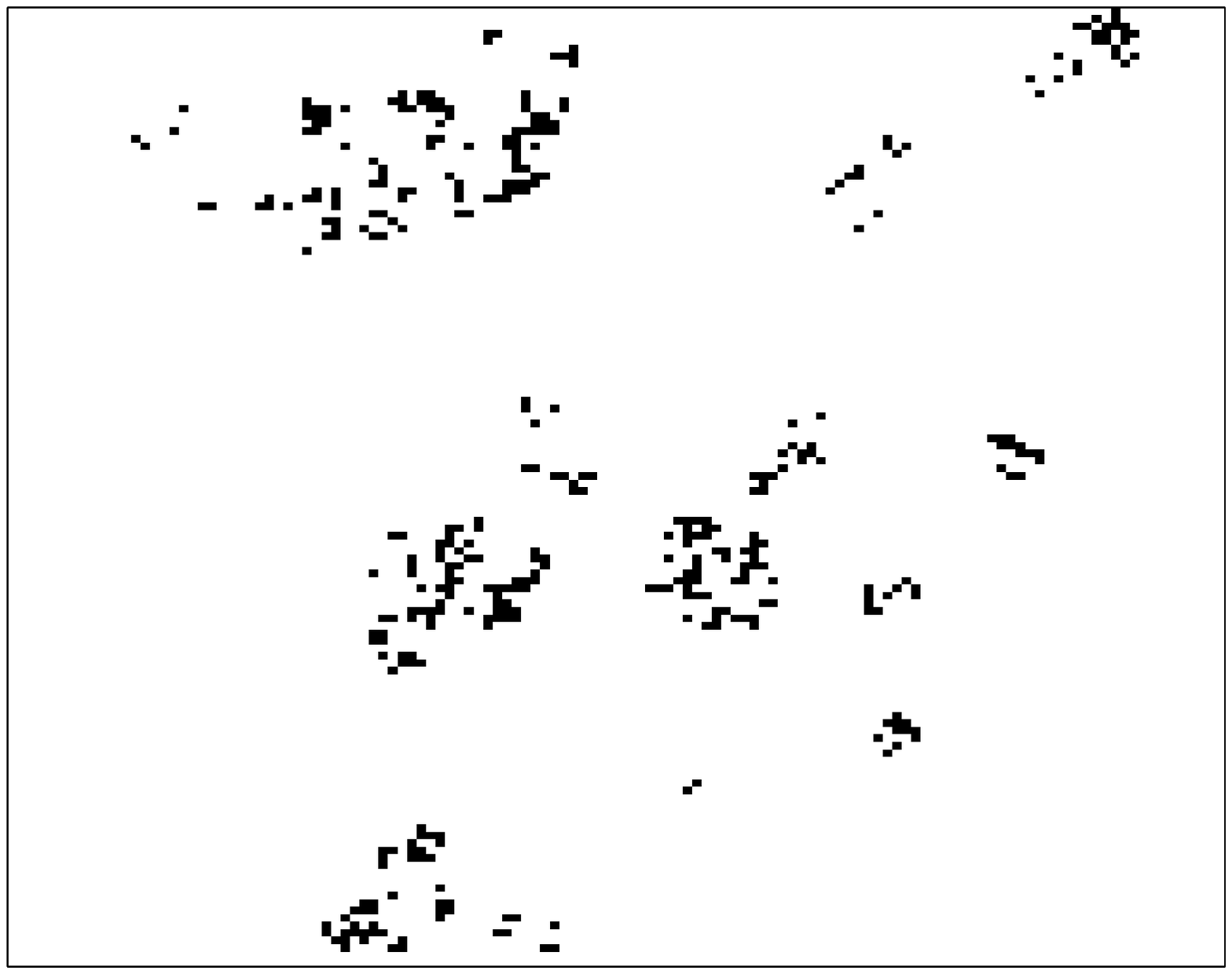}
        \caption{Typical plots for sexual reproduction only (left) and different
        rates of both sexual and asexual reproduction (middle and right).}
        \label{F: Snapshot examples}
        \end{figure}

\begin{figure}[tb]
        \centering\noindent
        \includegraphics[width=8.5cm]{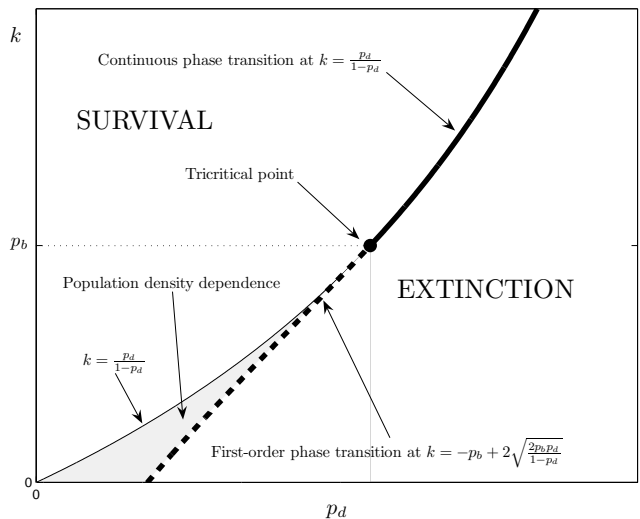}
        \caption{Phase diagram showing the continuous phase transition occurring
        for $k \ge\pb$ and the first-order phase transition for $k<\pb$.
        Eqns. (\ref{E: Critical Point 1}-\ref{E: Critical Point 2}) have         been re-arranged in the figure to make $k$ the dependent variable.}
        \label{F: Phase Diagram}
        \end{figure}

Following on from L\"ubeck \cite{Lubeck_Tricritical}, we introduce a lower-order proliferation reaction to our original model. By allowing both asexual and sexual reproduction, we have a phase space exhibiting first-order, continuous and tricritical behavior. Whilst the dynamics of such a system have been previously studied \cite{Lubeck_Tricritical, Grassberger_Tricritical},
the geometrical structure of the population at the different types of critical
behavior has so far been ignored. Sample snapshots of the population for
sexual reproduction only and both asexual and sexual reproduction are shown
in Fig. \ref{F: Snapshot examples}. In this letter, we study this geometrical
structure and show the remarkable result that the relationship between
initial cluster size and the survival probability up to some finite time
$t$ is linear at the tricritical point only.
\\ \\ 
\textit{The Model: }
We have a $d$-dimensional square lattice of linear length $L$ where each site is either
occupied by a single particle or is empty. A site is chosen at random. The particle on an occupied site dies with probability $p_{\rm d}$,  leaving the site empty. If the particle does not die, a nearest neighbor site is randomly chosen. If the neighboring site is empty the particle moves there and produces a new individual at the site that it has just left with probability $k$. If the chosen site is however
occupied, the particle reproduces with probability $p_{\rm b}$ producing a new particle on another randomly selected neighboring site, conditional on that site being empty. A time step is defined as the number of lattice sites $N=L^{\rm d}$ and periodic boundary conditions are used. 
We have the following reactions for a particle $A$ for proliferation and
annihilation respectively,
        \begin{equation} \label{E: Reactions}
        A+A+\phi\longrightarrow3A, \quad A+\phi\longrightarrow 2A \quad \mbox{and}         \quad A\longrightarrow \phi.
        \end{equation}
Assuming the particles are spaced homogeneously, the mean field equation
for the density of active sites $\rho(t)$ is given by:
        \begin{eqnarray} \label{E: Mean Field}
        \pderiv{\rho(t)}{t} & = & \pb\left(1-\pd\right)\rho(t)^2\left(1-\rho(t)\right)
        \nonumber \\
        & + & k(1-\pd)\rho(t)\left(1-\rho(t)\right)-\pd\rho(t).
        \end{eqnarray}
The first term considers the sexual reproduction term, the second asexual
reproduction, and the final term death of an individual. Eq. \ref{E: Mean
Field} has three stationary states:
        \begin{eqnarray} \label{E: Steady States}
        \bar\rho_0 & = & 0,\\
        \bar\rho_\pm & =& \frac{1}{2}
        \left[1-\frac{k}{\pb}\pm\sqrt{\left(\frac{k}{\pb}-1\right)^2
        +\frac{4}{\pb}\left(k-\frac{\pd}{1-\pd}\right)}\right]. \nonumber
        \end{eqnarray}  
For $k \ge \pb$, $\rhobar_+\rightarrow 0$ continuously as $\pd \rightarrow k/(1+k)$, indicative of a continuous phase transition with critical point
        \begin{equation} \label{E: Critical Point 1}
        \pdc = \frac{k}{1+k}.
        \end{equation}
For $k < \pb$, however, we have a jump in $\rhobar_+$ from $(\pb-k)/2\pb$ to
zero, this time at the critical point
        \begin{equation} \label{E: Critical Point 2}
        \pdc = \frac{(k+\pb)^2}{4\pb +(k+\pb)^2}.        
        \end{equation}
Further, for $k < \pb$, we have a region 
        \begin{equation}
        \frac{k}{1+k} \;<\; \pd \;\leq\; \frac{(k+\pb)^2}{4\pb +(k+\pb)^2}
        \end{equation}        
where the survival of the population is dependent on the population density. In fact, we have extinction for
        \begin{equation}
        \rho(t) < \rhobar_-(k < \pb, \pd)
        \end{equation} 
since then $\partial \rho/\partial t < 0$. For $k < \pb$ we therefore have a first-order phase transition. The two lines
of phase transitions meet at the point $k = \pb$, defining the position of
the tricritical point $k^*$. At the mean field level then, we have a phase
diagram as shown in Fig. \ref{F: Phase Diagram}.
We note that in our earlier paper \cite{Windus}, we examined the case $k=0$ and so were restricted, at the mean field level, to the first-order regime only. 

From simulations we find that for 1+1 dimensions, continuous phase transitions
are observed for all $k$. In order to examine tricritical behavior we therefore
proceed in 2+1 dimensions.

\begin{figure}[tb]
        \centering\noindent
        \includegraphics[width=8cm]{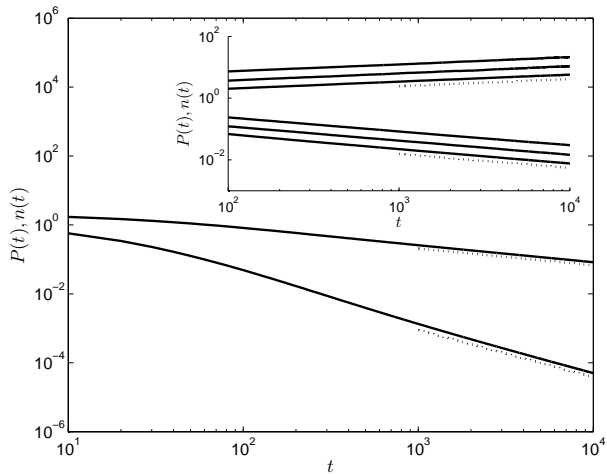}
        \caption{Plots showing the power-law behavior for $n(t)$ and
        $P(t)$ (bottom) outlined in Eqns. (\ref{E: eta}-\ref{E: delta'}).
        The main plot is with $k = k^* = 0.12$ and $\pd=\pdc = 0.1502195$.         We find $\eta = -0.487(5)$ and $\delta'=1.36(5)$. The inset shows         (from top to bottom)
        $k = 0.5$, 0.75 and 1. The values of $\pdc$ are               found to be 0.30302(0), 0.36765(0) and 0.41626(0) respectively. The               hashed lines show the power-law with DP values.}
        \label{F: Power-Law}
        \end{figure}

To examine the critical behavior of our model, we study the dynamical
behavior of the system after starting from a single-seed at the centre of
a large lattice. The lattice is sufficiently
large so that no individuals reach the boundary during the running of the
simulation. For fixed $\pb = 0.5$ we find the critical value
$\pdc$ for a given $k$. For the continuous phase transitions, at $\pd = \pdc$,
we expect power-law behavior
        \begin{eqnarray}
        n(t) & \;\bar\propto\; & t^\eta \label{E: eta}\\
        P(t) & \;\bar\propto\; & t^{-\delta'} \label{E: delta'} 
        \end{eqnarray}
for the population size and probability of survival respectively. For $k$
sufficiently greater than $k^*$, we find DP values $\eta = 0.231$ and $\delta' = 0.451$ \cite{Ziff} as shown in the inset of Fig. \ref{F: Power-Law}.
At $k = k^*$ we again expect power-law behavior because of the continuous
phase transition, but with different values for the exponents \cite{Ohtsuki1,Ohtsuki2}.
The best power-law is found for $k=k^*= 0.12(0)$ with $\pd = \pdc = 0.150219(5)$
(see Fig. \ref{F: Power-Law}), where the numbers in the perenthesis show the uncertainty in the last figure. For more details of methods obtaining the tricritical point, see Refs. \cite{Lubeck_Tricritical, Grassberger_Tricritical}.

To find the values of $\pdc$ in the first-order regime we adopt a different
approach, inspired by Lee and Kosterlitz \cite{Lee_New}, due to the lack of power law behavior.
For $k < k^*$ we examine histograms of population density against frequency
for different values of $\pd$. Due to the phase-coexistence at the critical point, we observe a double-peaked structure in the histogram for $\pd$ close to $\pdc$, where the peaks are at equal heights for $\pd = \pdc$. 
\\ \\
\textit{Cluster Geometry: } 
The difference between the mean field and simulation results is due to both
the neglect of the fluctuations in Eqn. (\ref{E: Mean Field}) and the false assumption
of a homogenous distribution. In this letter, we are particularly interested in the heterogeneous distribution and in particular the geometrical structure of the clusters at criticality and how this changes with $k$.
        \begin{figure}[tb]
        \centering\noindent
        {\small a)} \\
        \includegraphics[width=1.8cm,height=1.8cm]{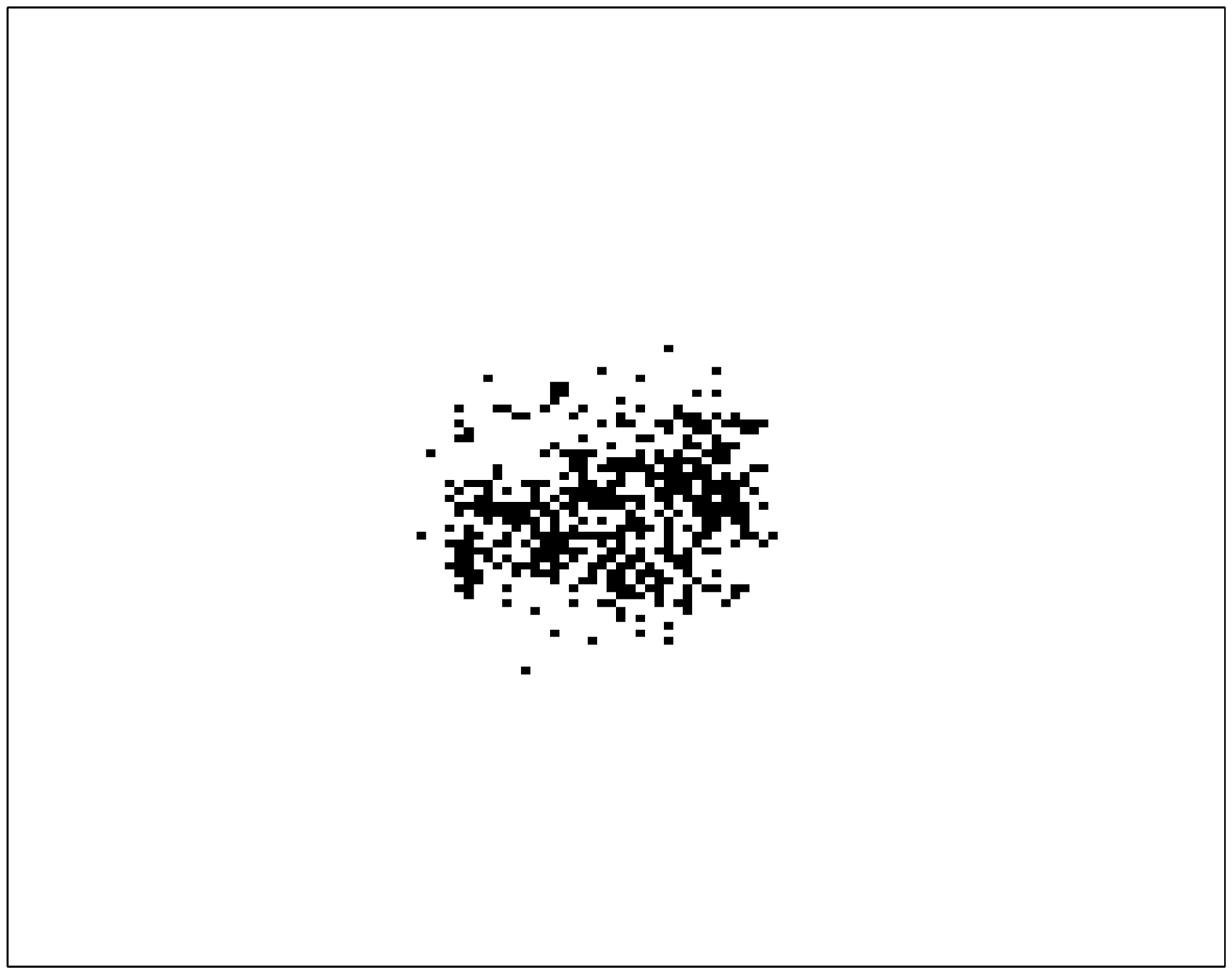}
        \includegraphics[width=1.8cm,height=1.8cm]{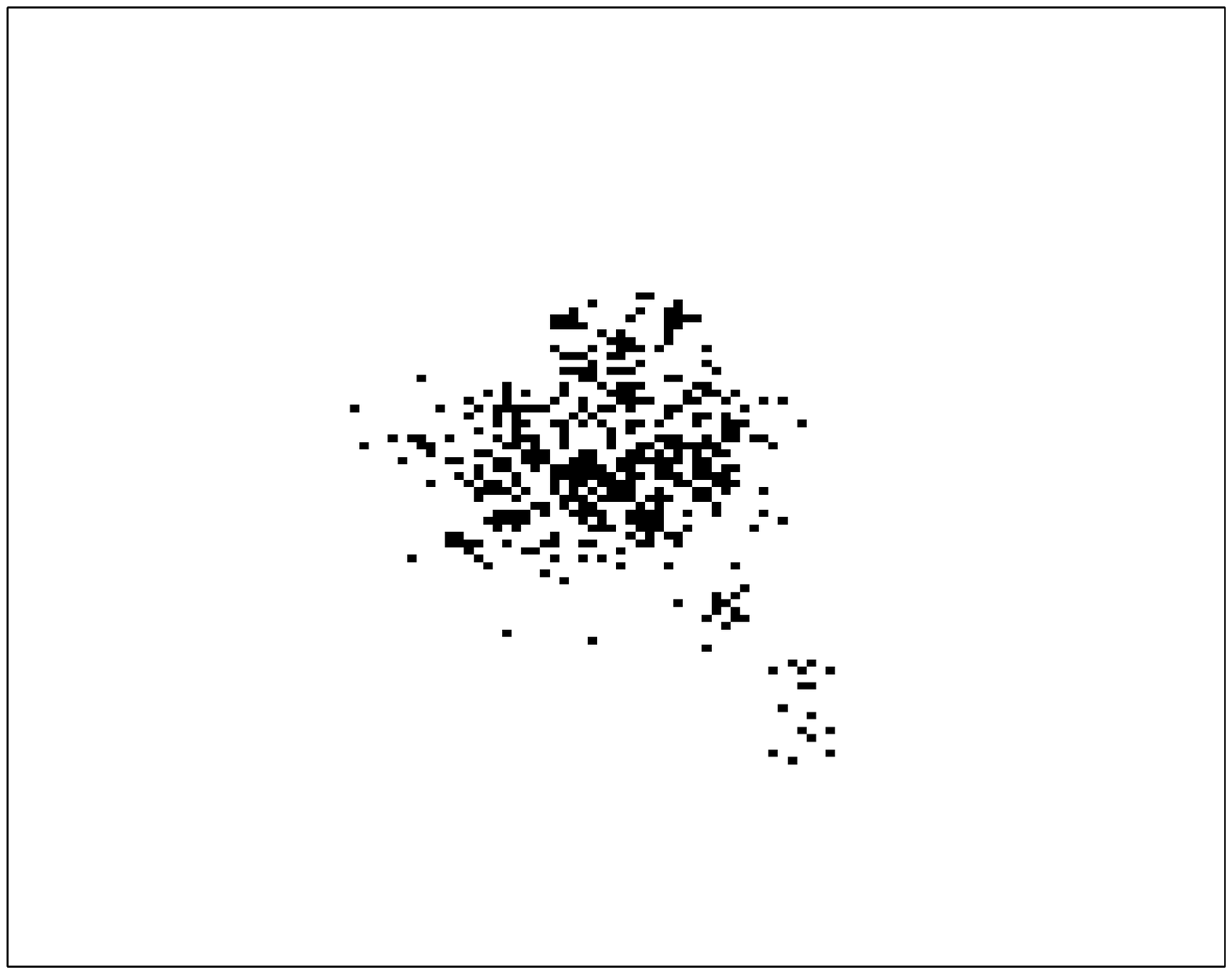}
        \includegraphics[width=1.8cm,height=1.8cm]{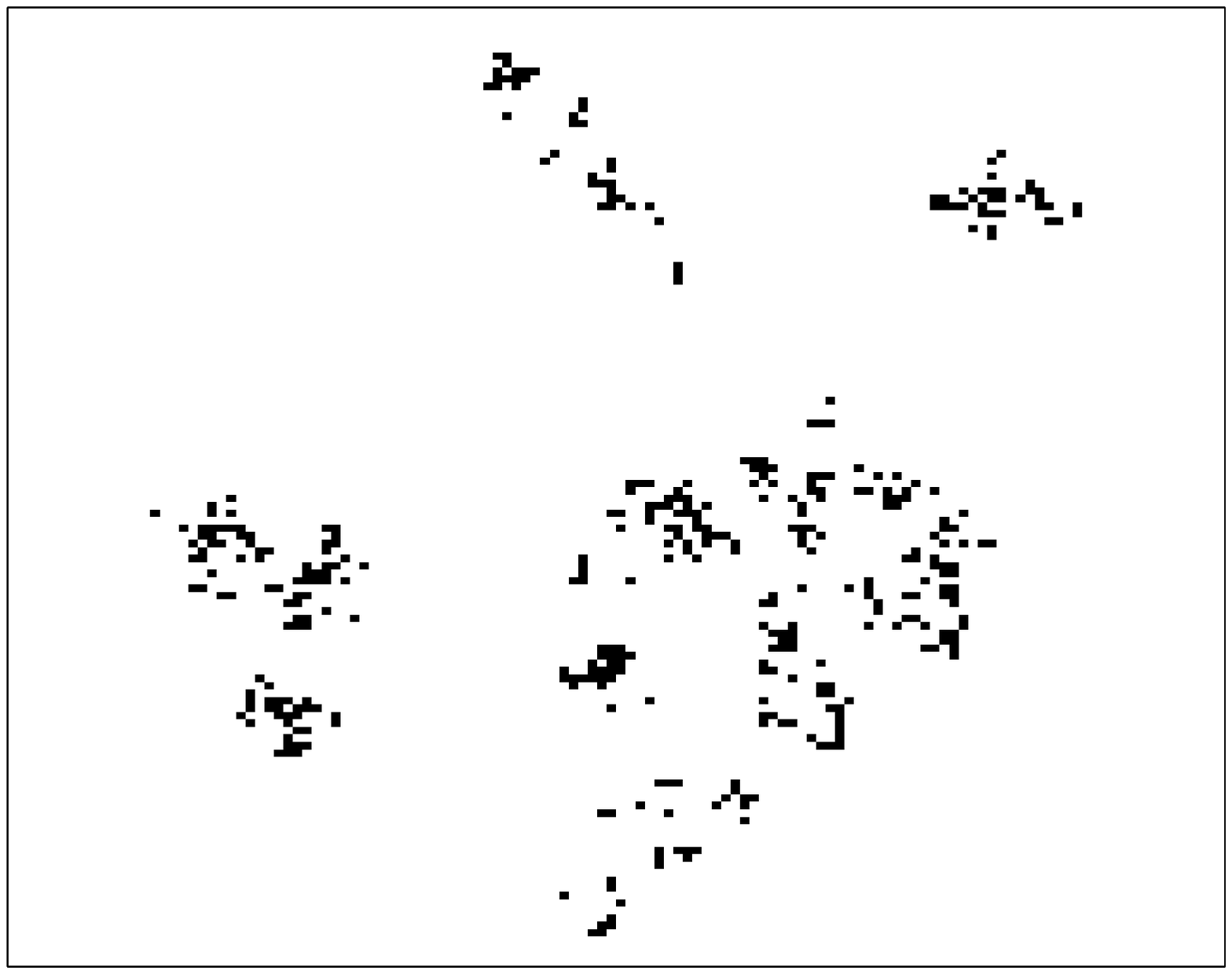}
        \includegraphics[width=1.8cm,height=1.8cm]{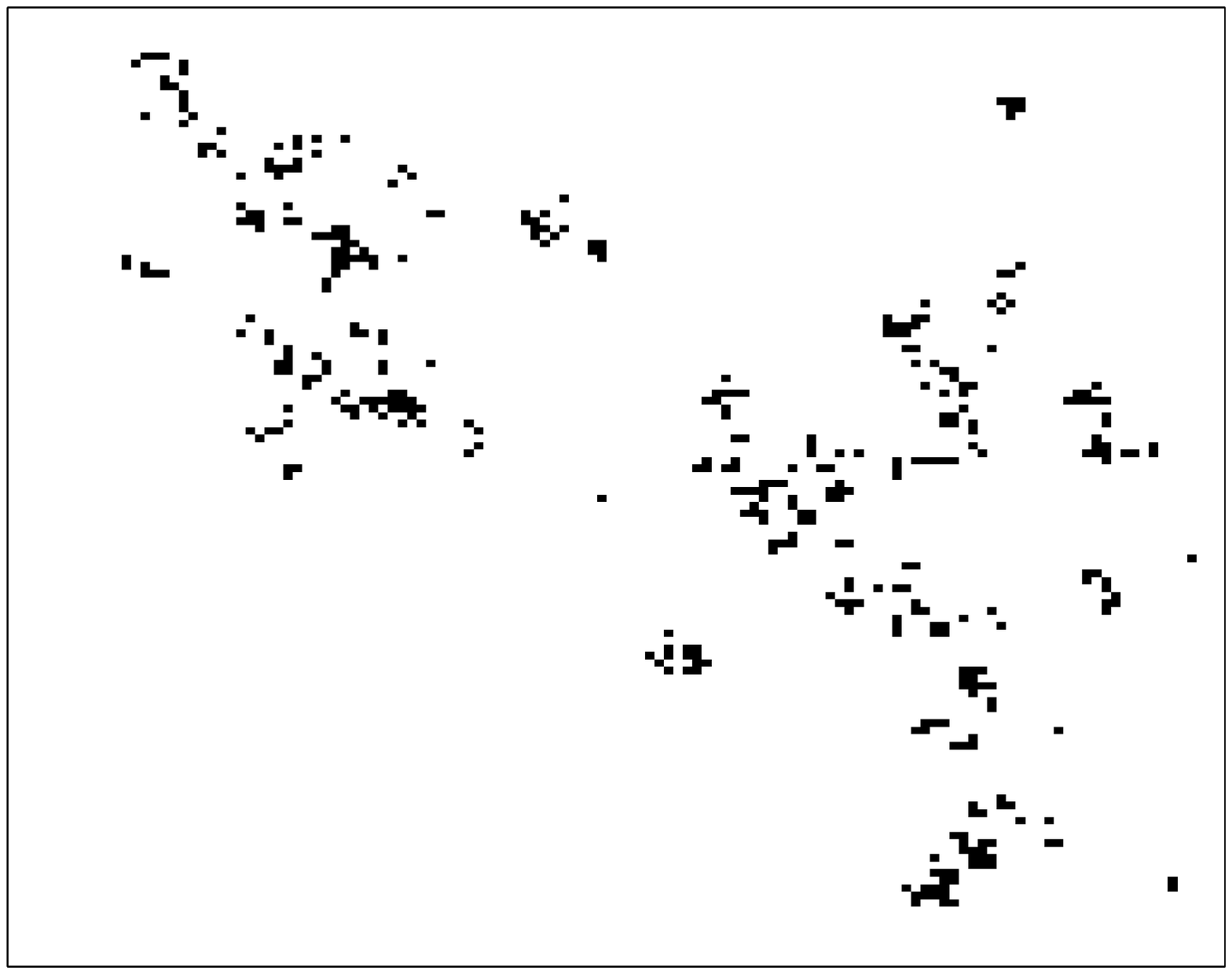} \\
        {\small b)} \\
        \includegraphics[width=1.8cm,height=1.8cm]{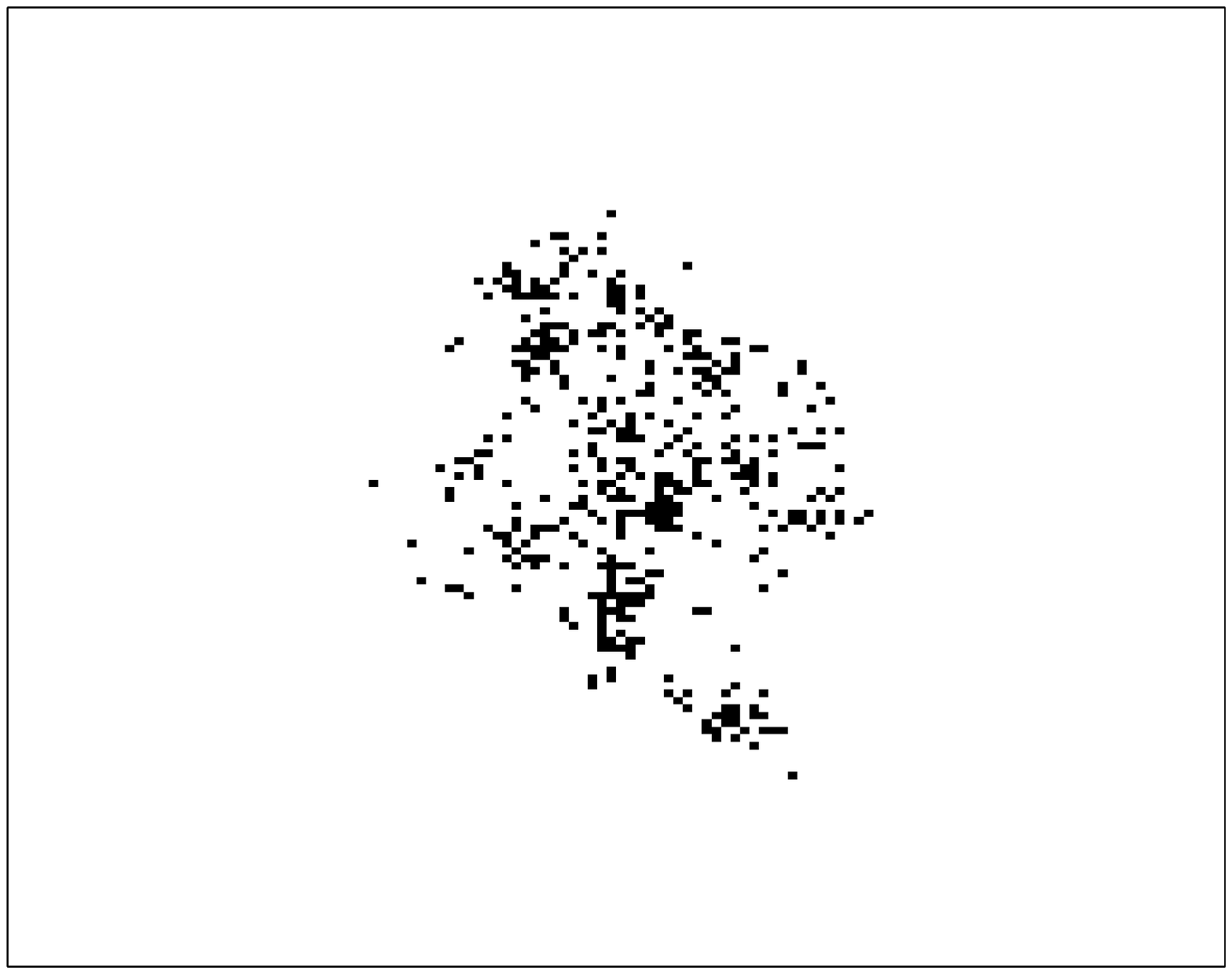}
        \includegraphics[width=1.8cm,height=1.8cm]{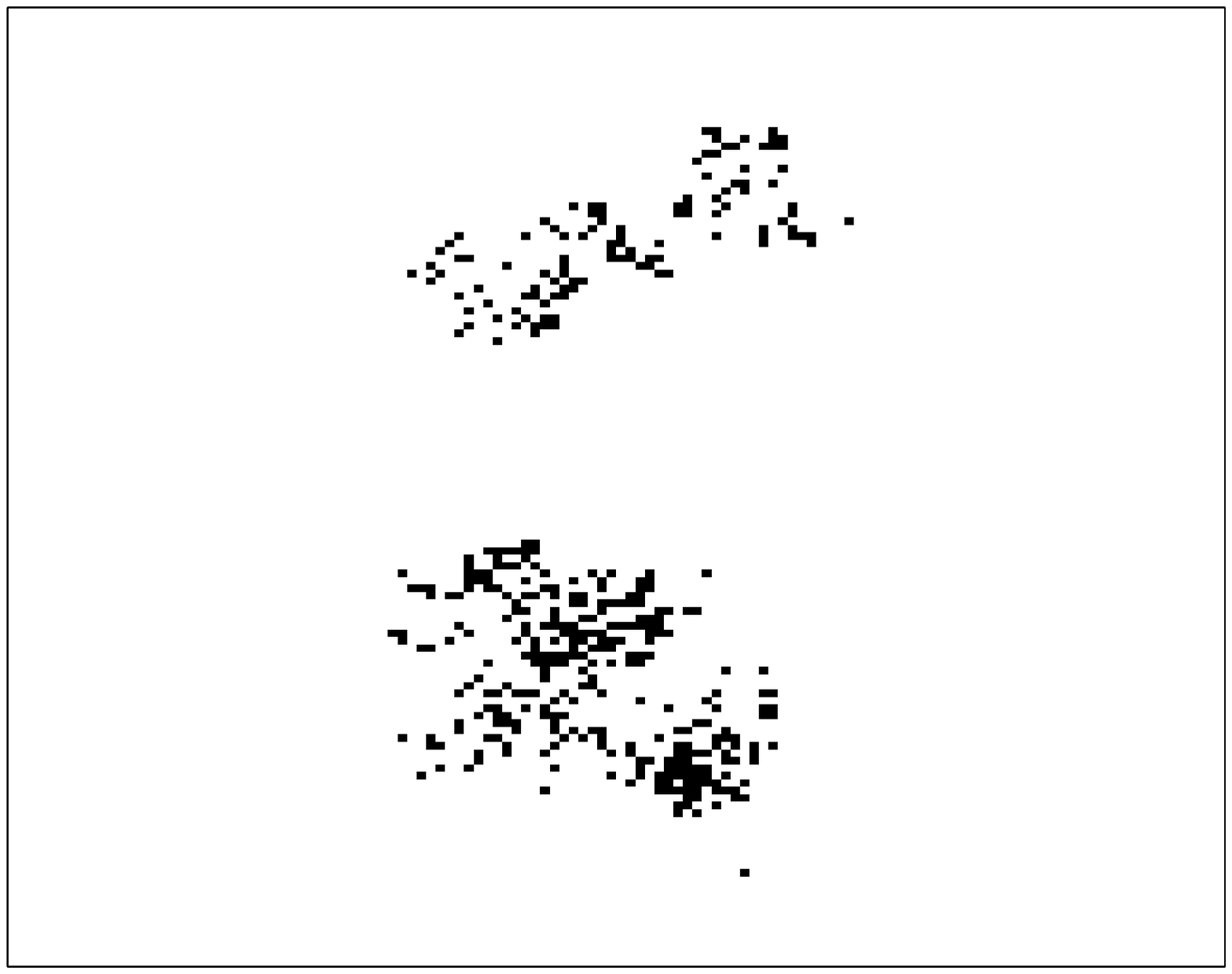}
        \includegraphics[width=1.8cm,height=1.8cm]{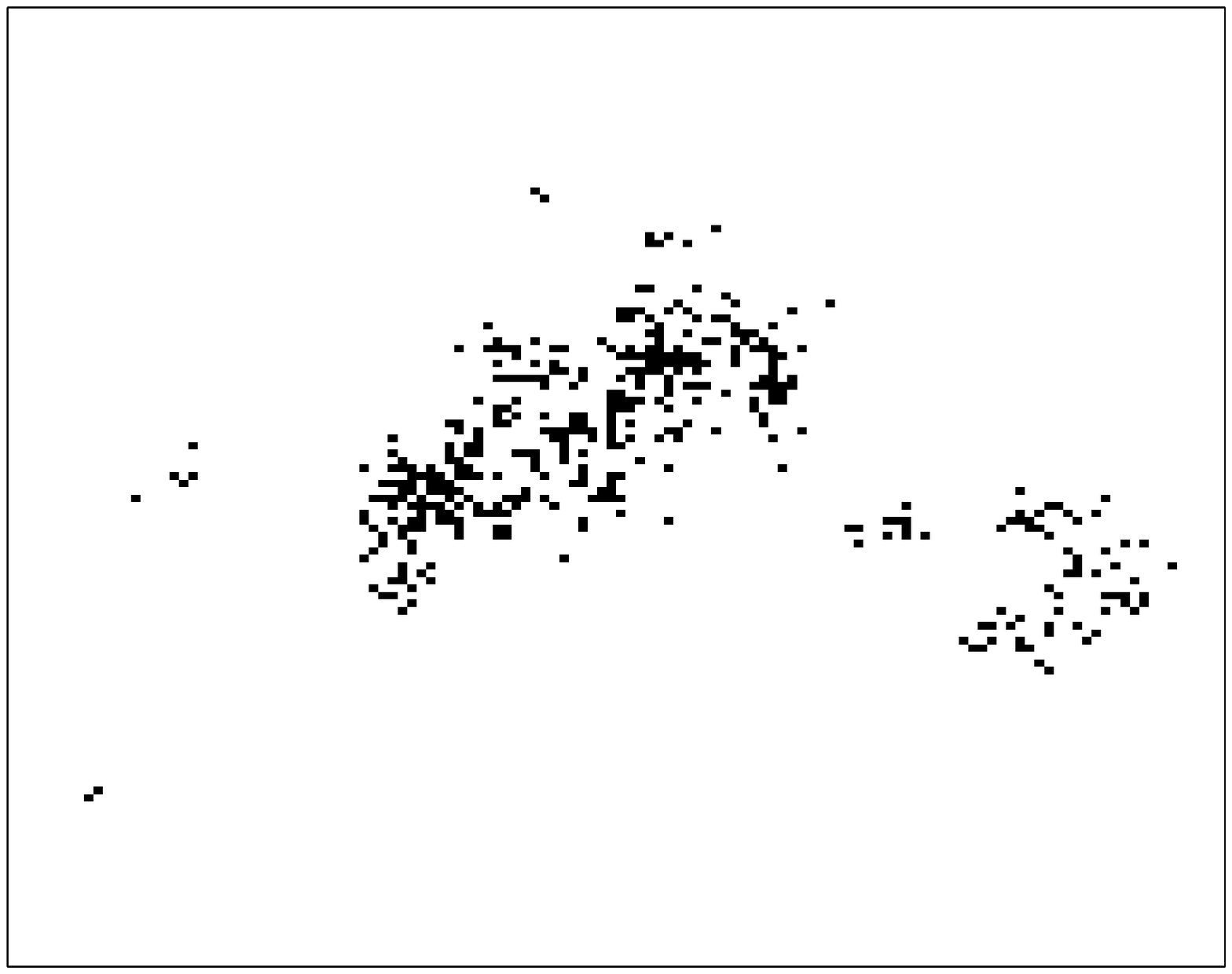}
        \includegraphics[width=1.8cm,height=1.8cm]{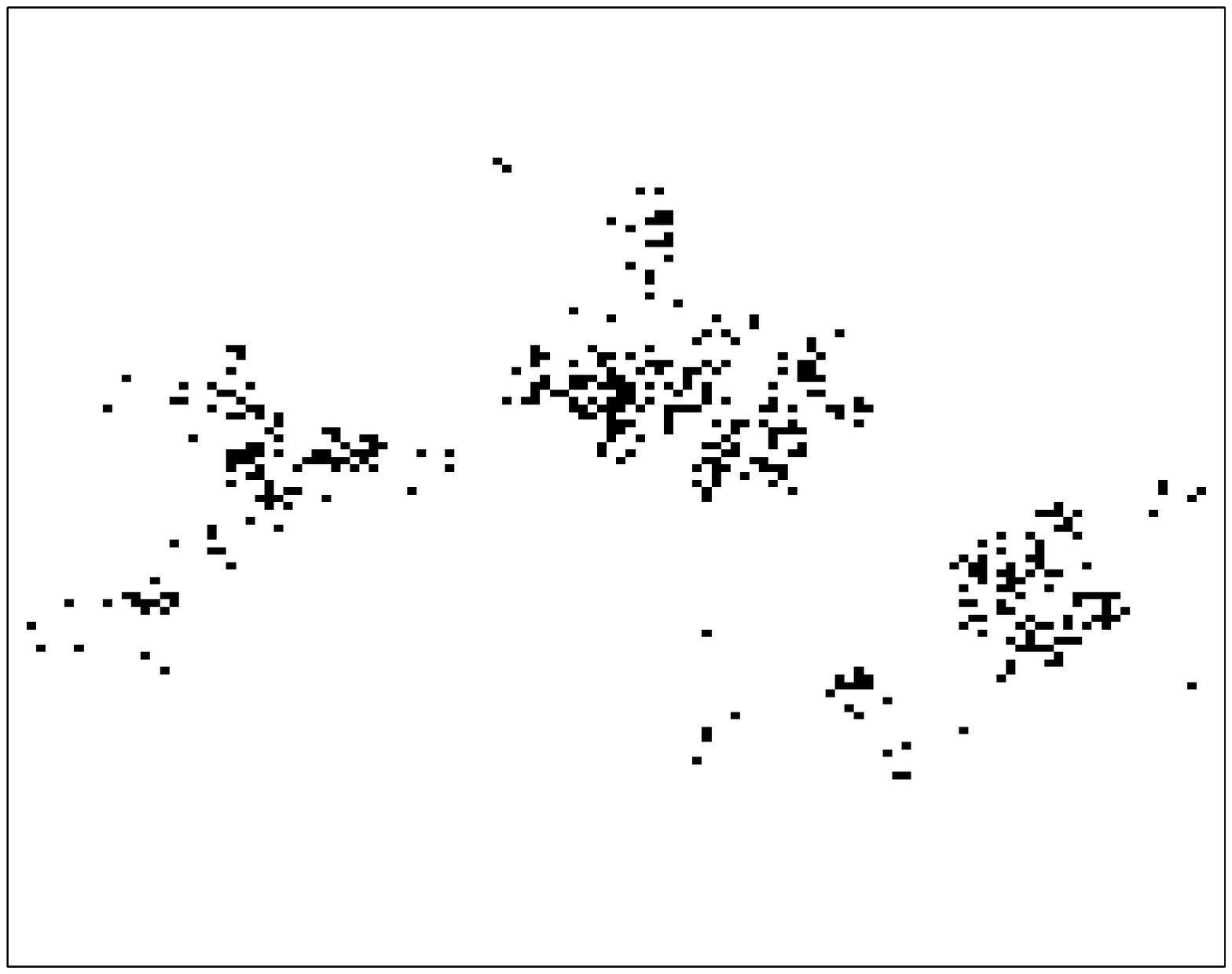}
        \caption{Typical plots of the population structure at the critical
        point for a) (from left
        to right) $k = 0$, 0.05, 0.5 and 1 and b) $k=k^*=0.12$. The simulations
        began from an initially fully occupied $128\times128$ lattice and
        ran until the population size reached 400 for the first time.}
        \label{F: Snapshots}
        \end{figure}
From Fig. \ref{F: Snapshots} a), we see that for $k \ll k^*$,
after starting from a fully occupied lattice, the population becomes densely clustered. This is certainly not unexpected
since, for $k = 0$, reproduction is only possible if an individual finds another. In contrast, in the DP region, the population is much more dispersed. For $k\gg k^*$, an individual has a greater chance of reproduction
if it is surrounded by empty sites and so being heavily clustered is a disadvantage.
For $k \neq k^*$, the plots shown in Fig.
\ref{F: Snapshots} a) are very typical plots, whereas at the tricritical
point, the plots are very varied. For $k =k^*$, Fig. \ref{F: Snapshots} b) shows that the population has little preference as to its structure. There are often areas where the population
is heavily clustered as well as those where the structure is DP-like. 

\begin{figure}[tb]
        \centering\noindent
        \includegraphics[width=8cm]{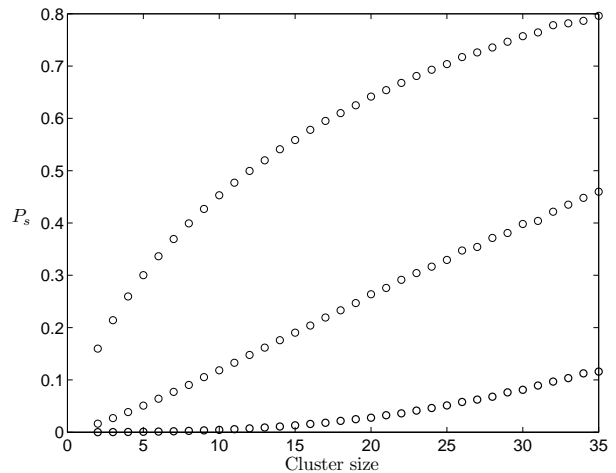}
        \caption{Plot of survival probability (for $t_{\rm max} = 200$),         against initial cluster size
        for (from bottom to top) $k = 0$, 0.12 and 1. Relatively small cluster
        sizes were used since clearly $P_{\rm s}\rightarrow 1$ as cluster
        size $\rightarrow\infty$.}
        \label{F: Cluster Size}
        \end{figure}

In considering the geometrical make-up of the clusters, it is interesting
to examine how the survival probability of a cluster depends on its size
for different values of $k$. To examine this, we run simulations from an
initially fully-occupied lattice up to the point when a certain population
is reached. We ensure that this population is sufficiently small so that
the resulting density does not `force' the population to take on certain cluster sizes. Once this population has been reached, we make a copy of each cluster
and place each one in the centre of a sufficiently large lattice. The survival probability $P_{\rm s}$ up to some time $t_{\rm max}$ of the population is then measured for each individual initial cluster. The results for each cluster size are shown in Fig. \ref{F: Cluster Size}.
        
\begin{figure}[tb]
        \centering\noindent
        \includegraphics[width=8cm]{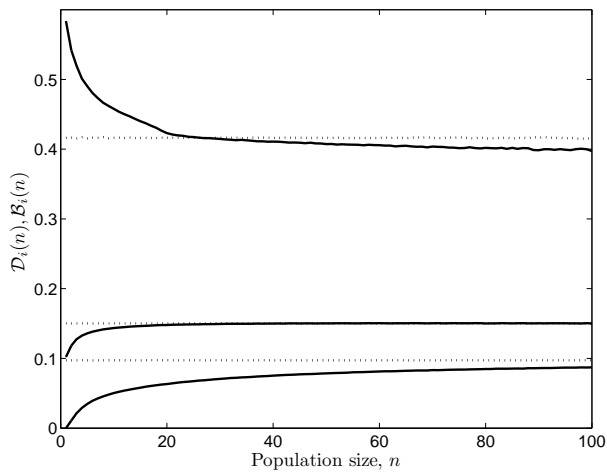}
        \caption{The number of births (solid line) and deaths (hashed line)         per individual per time step for (from top to bottom) $k = 1$, 0.12
        and 0.}
        \label{F: BirthsDeaths}
        \end{figure}

We see that the gradient of $P_{\rm s}$ with respect to cluster
size increases for $k = 0$ whilst it decreases for $k = 1$. Consistent with
the plots from Fig. \ref{F: Snapshots}, we have that additional individuals
give an increasing advantage to the survival probability for $k=0$ but
a negative effect for $k = 1$. Due to the curvature of the plots for $k = 1$ and $k=0$, we would expect a linear relationship for some $0 < k < 1$. Surprisingly this occurs at the tricritical point $k = k^* = 0.12$. In order to understand why this linear behavior might occur
at the tricritical point, we plot in Fig. \ref{F: BirthsDeaths} the average number of births and deaths that occur per individual per time step, $\Bi(n)$
and $\Di(n)$ respectively, at the critical point for different $k$.
As expected, for all values of $k$, $\Di(n)$ remains constant for all population sizes. In contrast, an increasing
population seems to have a positive influence on $\Bi(n)$ for
$k = 0$ whilst a negative influence for $k = 1$. At $k = k^*$, for sufficiently
large population sizes, $\Bi(n)$ is independent
of the population size. In fact, for $n > 40$, $\Bi(n) \approx
\Di(n)$.

We examine now whether it is this equality in the number of births and deaths
that results in the observed linear relationship between initial cluster size and
$P_s$. We simplify our model by considering the macroscopic population size $n(t)$ only and ignoring the microscopic individuals. Analogous to a random walk, the population increases
by one with probability $\pb$, decreases by one with probability $\pd$ and
stays the same with probability $1-\pb-\pd$. We again examine the probability
that the population survives up to some time $t_{\rm max}$ where a time step
at time $t$ is defined as $n(t)$ updates. We note that this is the same definition
of a time step that we use when running simulations that begin with a single-seed.
In the simulations, it corresponds to approximately one update per individual.

With $\pb = \pd$, we find that
the probability of survival increases linearly with initial population size.
This finding seems to indicate the equality in the number of births/deaths
per individual per time step results in the observed linear behavior at
the tricritical point.

\textit{Conclusion: }We have examined the critical behavior in our model which exhibited continuous,
first-order and tricritical phase transitions. We noted the key geometrical
differences in the cluster structure for each type of transition caused by
the different values of the rate of asexual reproduction $k$. Of particular
importance was the surprising result of the linear increase in probability
of survival with cluster size. We hypothesised that this was due to the number
of births and deaths per time step being equal at the tricritical point.

The lack of sensitivity of this linear behavior to the parameters do not
make this an effective method to obtain the position of the tricritical point.
To do this, we believe seeking the power law behavior for $n(t)$ and $P(t)$
to be the best approach.

One area of future study is the discrepency between our values for $\eta$
and $\delta'$ at the tricritical point with Grassberger's 
who found $\eta = -0.353(9)$ and $\delta'=1.218(7)$ \cite{Grassberger_Tricritical}.
Indeed, he had disagreement between his values for $\beta$, $\nu_\perp$ and
$\phi$ and L\"ubeck's \cite{Lubeck_Tricritical}. Universality at the
tricritical point seems to have been assumed but the differences in these
critical exponents appear to dispute that. Further study needs to be
done to clarify the situation.
 
We are indebted to Sven L\"ubeck whose correspondence to us about our first paper on this subject led us to this research. All computer simulations were carried out on the Imperial College London's HPC. Alastair Windus would also like to thank EPSRC for his Ph.D. studentship.


\end{document}